\documentstyle[aps,prl,floats,epsf,epsfig,amssymb]{revtex}

\newcommand{\bastar}{\begin{eqnarray*}}
\newcommand{\eastar}{\end{eqnarray*}}
\newskip\humongous \humongous=0pt plus 1000pt minus 1000pt

\newif\ifdtup

\relax

\draft 
\preprint{SNUTP 02/002}
\begin{document}
\twocolumn[\hsize\textwidth\columnwidth\hsize\csname@twocolumnfalse%
\endcsname
\title{\Large\bf Dual Description of Brane World
Cosmological Constant with $H_{MNPQ}$ }
\bigskip

\author{Kang-Sin Choi,
Jihn E. Kim
and Hyun Min Lee}
\address{Department of Physics and Center for Theoretical
Physics, Seoul National University, Seoul 151-747, Korea \\
\vskip 0.3cm
}
\maketitle

\begin{abstract}
We present a short review of the recent 5D self-tuning solution of
the cosmological constant problem with $1/H^2$ term, and present
the dual description of the solution. In the dual description, we
show that the presence of the coupling of the dual field($\sigma$)
to the brane(which is a bit different from the original theory)
maintains the self-tuning property.

\vspace{0.3cm}
\end{abstract}

\narrowtext
\bigskip
                        ]
\def\L{$\Lambda$}
\def\F{$\{F\tilde F\}$}
\def\p{\partial}

\section{Introduction}

The cosmological constant problem is the most severe hierarchy
problem or fine-tuning problem known to particle physicists since
1975\cite{veltman,weinberg}. It was introduced in the Einstein
equation as
\begin{equation}
R_{\mu\nu}-\frac{1}{2}Rg_{\mu\nu}-8\pi GV_0g_{\mu\nu}=8\pi
GT_{\mu\nu}\label{ehleq}
\end{equation}
where the term $8\pi GV_0$($V_0$ is the vacuum energy density
considered in particle physics.) is the so-called cosmological
constant(c.c.) $\Lambda$. One can see that it is quite natural to
introduce a constant in the action.  Therefore, the constant is of
order the mass scale in question. The parameter appearing in
gravity is the Planck mass $M=2.44\times 10^{18}$~GeV which is
astronomically larger than the electroweak scale. Since gravity
introduces a large mass $M$, any other parameter in gravity is
expected to be of that order, which is a natural expectation.
Namely, $V_0$ is expected to be of order $M$. However, the bound
on the vacuum energy was known to be $<(0.01\ {\rm eV})^4$, which
implied a fine-tuning of order $10^{-120}$.

This c.c. problem surfaced as a very serious one when one
considered the spontaneous symmetry breaking in particle
physics~\cite{veltman}. The minimum position of potential does not
matter in particle physics. But in gravity its position determines
the c.c. There have been several attempts toward the solution, by
Hawking\cite{hawking}, Witten\cite{witten},
Weinberg\cite{anthropic}, Coleman\cite{coleman}, etc, under the
name of probabilistic interpretation in Euclidian gravity,
boundary of different phases, anthropic solution, wormhole
solution, etc.

The self-tuning solution, which attracted a great deal of
attention recently, is basically that the equations of motion
choose the vanishing cosmological constant solution. For example,
if the flat space solution exists, then it is a zero cosmological
constant solution. In 4D, a nonzero vacuum energy does not allow
the flat space solution. Thus, to obtain the flat space one has to
fine-tune $V_0$ to zero very accurately.

In Sec. II, we review the self-tuning ideas. In Sec. III, we
summarize the self-tuning solution with $H_{MNPQ}$ discussed in
Ref.\cite{kklcc0,kklcc}. In Sec. IV, we present a self-tuning
solution in the dual description in terms of $\sigma$. This
self-tuning property is shown to be maintained even though we
introduce the brane coupling of the dual field $\sigma$. Sec. V is
a conclusion.

\section{Self-tuning ideas}

We can distinguish the self-tuning solutions into the old version
and the new version.
\subsection{Old version}
If there exists a solution for the flat space, then the existence
itself is called a self-tuning solution, or a solution with the
undetermined integration constant(UIC).  For a nonzero $\Lambda$
in 4D, a flat space ansatz $ds^2=d{\bf x}^2-dt^2$ does not allow a
solution. The de Sitter space(dS, $\Lambda>0$) or anti de Sitter
space(AdS, $\Lambda<0$) solution is possible. To reach a nearly
flat space solution, one needs an extreme fine tuning, which is
the c.c. problem.

But suppose that there exists an UIC. Witten used the four index
field strength  $H_{\mu\nu\rho\sigma}$ to obtain an
UIC\cite{witten}. The equation of motion of $H$ leads to an UIC,
say $c$. Thus, the vacuum energy contains a piece $\sim c^2$. This
UIC $c$ can be adjusted so that the final c.c. is zero. Once $c$
is determined, there is no more UIC because $H_{\mu\nu\rho\sigma}$
is not a dynamical field in 4D. When vacuum energy is added later,
there is no handle to adjust further. In a sense, it was another
way of fine-tuning. However, if there exists a dynamical field
allowing an UIC, it is a desired old style self-tuning solution.
This old version did not care whether there also exist de Sitter
or anti de Sitter solutions. Selection of the flat space out of
these solutions is from a principle such as Hawking's
probabilistic choice.
\subsection{New version}
In recent years, a more ambitious attempt was proposed, where only
the flat space ansatz has the solution\cite{kachru}.(But at
present there does not exists a solution in this category.) The
Randall-Sundrum II type models\cite{rs2} were constructed in 5D
bulk AdS, i.e. the 5D bulk cosmological constant $\Lambda_b<0$,
with brane(s) located at the fixed point at $y=0$. The RS II model
uses only one brane. At this brane one can introduce a brane
tension $\Lambda_1$. Thus, the gravity Lagrangian contains two
free parameters $\Lambda_b$ and $\Lambda_1$. In Fig. 1, we show
this situation schematically, where the extra dimension is the
$y$-direction. $x^\mu$ is the 4D index. The 4D flat space ansatz
allows a solution for a specific choice of $k_1$(basically
$\Lambda_1$, $k_1=\Lambda_1/6M^3$) and $k$(basically $\Lambda_b$,
$k=\sqrt{-\Lambda_b/6M^3}$): $k_1=k$. Therefore, it requires a
fine-tuning between parameters as in the 4D case. However, it
seems to be an improvement since we reach at the 4D flat space
from nonzero cosmological constants and the RS II type models
seems to be a good playground to obtain UIC solutions.

The RS II model is an interesting extension of the space-time
without compactification. With the bulk AdS, the uncompactified
fifth dimension can be acceptable due to the localized
gravity\cite{rs2}. For the brane located at $y=0$, the action is
\begin{eqnarray}
S&=&\int d^4x \int dy \sqrt{-g}\bigg(
\frac{M^3}{2}(R-\Lambda_b) \nonumber \\
&+&(-\Lambda_1+{\cal L}_{matter})\delta(y) \bigg)
\end{eqnarray}
where $M$ is the 5D fundamental scale and ${\cal L}_{matter}$ is
the matter Lagrangian, assuming the matter is located at the brane
only. The flat space ansatz,
\begin{equation}
ds^2=\beta(y)^2\eta_{\mu\nu}dx^\mu dx^\nu+dy^2\label{flat}
\end{equation}
allows the solution if $k=k_1$. Even though the 4D is flat, it is
curved in the direction of the fifth dimension, denoted by the
warp factor $\beta(y)=\beta_0\exp(-k|y|)$. Namely, the gravity is
exponentially unimportant if it is separated from the brane. [If
there were more branes, there are more conditions to satisfy
toward a flat space since one can introduce a brane tension at
each brane. Thus, the RS II type models are the simplest ones.]

The try to obtain a new type of self-tuning solution was initiated
by Arkani-Hamed et. al. and Kachru et. al.\cite{kachru}. For
example the 5D Lagrangian with the introduction of a massless bulk
scalar field $\phi$, coupling to the brane tension,
\begin{equation}
{\cal L}=R-\Lambda e^{a\phi}-\frac{4}{3}(\nabla\phi)^2-Ve^{b\phi}
\delta(y)\label{kachru}
\end{equation}
where we set $M^3/2=1$. We may ask, $\lq\lq$Why this Lagrangian?",
which involves more difficult related questions. Accepting this,
we must satisfy the following Einstein and field equations, with
the flat ansatz (\ref{flat}),
\begin{eqnarray}
({\rm dilaton})&:&
\frac{8}{3}\phi^{\prime\prime}+\frac{32}{3}A'\phi'-a\Lambda
e^{a\phi} -bV\delta(y)e^{b\phi}=0 \nonumber\\
(55)&:& 6(A')^2-\frac{2}{3}(\phi')^2+\frac{1}{2}\Lambda e^{a\phi}=0
  \nonumber\\
(55),(\mu\nu)&:&
3A^{\prime\prime}+\frac{4}{3}(\phi')^2+\frac{1}{2}e^{b\phi}V\delta(y)=0
\end{eqnarray}
where $2A(y)=\ln\beta(y)$, and prime denotes the derivative with
respect to $y$.

We will discuss the property of this solution somewhat in detail,
to compare with our new solution obtained in the dual picture.

For $\Lambda=0$, there exists a bulk solution satisfying
$A'=\alpha\phi'$,
\begin{equation}
\phi=\pm\frac{3}{4}\ln\left|\left(\frac{4}{3}\right)y+c\right|+d,\
\ \alpha=\pm\frac{1}{3}
\end{equation}
where $c$ and $d$ are determined without fine-tuning of the
parameters. The solution has a singularity at $y_c\equiv -(3/4)c$
or diverges logarithmically at large $|y|$. The logarithmically
diverging solution does not realize the localization of gravity.
If we restrict the space up to the singular point $y_c$, then at
every $y$ inside the space it is flat. However, the effective 4D
theory is the one after integrating out the allowed $y$ space.
Since $y_c$ is the naked singularity, we do not know how to cut
the $y$ integration near $y_c$, implying a possibility that the
flat space ansatz does not lead to a solution. Depending on how to
cut the integral, one may introduce a nonzero c.c. F$\ddot{\rm
o}$rste et. al. tried to understand this problem by curing the
singularity by putting a brane at $y_c$\cite{nilles}. Then, a flat
4D space solution is possible but one needs a fine-tuning. It is
easy to understand. If one more brane is introduced, then there is
one more tension parameter $\Lambda_2$, i.e. in the Lagrangian one
adds $\Lambda_2\delta(y-y_c)$. If the space is flat for one
specific value of $\Lambda_2$, then it must be curved for the
other values of $\Lambda_2$, since the $y$ integration gives a
c.c. contribution directly from $\Lambda_2$.

This example teaches us that the self-tuning solution better
should not have a singularity in the whole $y$ space. Even if we
neglect the problem of singularity, the brane interaction
Eq.(\ref{kachru}) must have {\it a specific value for $b$, which
is a fine-tuning.} Thus, we have not obtained a new self-tuning
solution yet.

\section{The self-tuning solution with $1/H^2$}

As we have seen in the RS II model, the Einstein-Hilbert action
alone does not produce a self-tuning solution. Inclusion of higher
order gravity does not improve this situation\cite{kklgb}. We need
matter field(s) in the bulk. The first try is a massless spin-0
field in the bulk as Ref.\cite{kachru} tried so that it affects
the whole region of the bulk. However, it may be better if there
appears a symmetry in the spin-0 sector. These are achieved by a
three index antisymmetric tensor field $A_{MNP}$. In 5D the dual
of its field strength is interpreted as a scalar. The field
strength $H_{MNPQ}$ is invariant under the gauge transformation
$A_{MNP}\rightarrow A_{MNP}+\partial_{[M}\lambda_{NP]}$, thus
masslessness arises from the symmetry. There will be one $U(1)$
gauge symmetry remaining with one massless pseudoscalar field
which is $a$, $\partial_M
a=(1/4!)\sqrt{-g}\epsilon_{MNPQR}H^{NPQR}$. But the interactions
are important for the solution, as in Ref.\cite{kachru} a bulk
solution was found for the specific form of the interaction.

The first guess is the bulk term $-(M/48)H^2$\cite{kklcc} where
$H^2\equiv H_{MNPQ}H^{MNPQ}$. The brane with tension $\Lambda_1$
is located at $y=0$, and the bulk c.c. is $\Lambda_b$. The ansatze
for the solution are
\begin{eqnarray}
{\rm Ansatz\ 1}\ &:&\ ds^2=\beta(y)^2\eta_{\mu\nu}dx^\mu
dx^\nu+dy^2,\nonumber \\ 
{\rm Ansatz\ 2}\ &:&\ H_{\mu\nu\rho\sigma}
=\epsilon_{\mu\nu\rho\sigma}\frac{\sqrt{-g}}{n(y)}\label{ansatz0}
\end{eqnarray}
where $\mu,\cdots$ are the 4D indices, and $n(y)$ is a function of
$y$ to be determined. It is sufficient to consider (55) and
($\mu\nu$) components Einstein equations and the $H$ field
equation. By setting $M=1$, we obtain the bulk
solution\cite{kklcc}
\begin{eqnarray}
\Lambda_b<0\ &:&\ \beta(|y|)=(\frac{a}{k})^{1/4}[\pm\sinh(4k|y|
+c)]^{1/4} \nonumber\\
\Lambda_b>0\ &:&\
\beta(|y|)=(\frac{a}{k})^{1/4}[\sin(4k|y|+c^\prime)
]^{1/4}  \nonumber\\
\Lambda_b=0\ &:&\
\beta(|y|)=|4a|y|+c^{\prime\prime}|^{1/4}.\label{h2}
\end{eqnarray}
For a localizable (near $y=0$) metric, there exists a singularity
at $y=-c/4k$, etc., except for some cases with $\Lambda_b>0$.
Thus, for these singular cases another brane is necessary to cure
the singularity, and we need a fine-tuning as in the case of
Kachru et. al.\cite{kachru}. The solution with the bulk de Sitter
space without a singularity, the second one in Eq. (\ref{h2}), is
worth commenting. Such a solution is periodic and depicted in Fig.
2. We can consider only $|y|\le y_c$, then $\beta^\prime=0$
at $y=\pm y_c$. The boundary condition at $y=0$ determines $c'=
\cot^{-1}(k_1/k)$ and the boundary condition at $y_c$ determines
$y_c$ such that $c'=4ky_c-\cot^{-1}(k_2/k)$, so it looks like an
UIC. Indeed, Ref.\cite{num} claims that such a solution is a
self-tuning one. But for $y_c$ to behave like an undetermined
integration constant, it should not appear in the equations of
motion. Note, however, that $y_c$ is the VEV of the radion
$g_{55}$, and hence it cannot be a strictly massless Goldstone
boson. If it were massless, it will serve to the long range
gravitational interaction and hence give different results from
the general relativity predictions in the light bending
experiments. Therefore, it should obtain a mass and hence $y_c$ is
not a free parameter but fixed. So the boundary condition at $y_c$
is a fine-tuning condition\cite{kklcc,radion}.

A working self-tuning model is obtained with the $1/H^2$
term\cite{kklcc0},
\begin{equation}
S=\int d^4x\int dy\sqrt{-g}\left( \frac{1}{2}R +\frac{2\cdot
4!}{H^2}-\Lambda_b-\Lambda_1\delta(y) \right).\label{action}
\end{equation}

\subsection{Flat space solution}

For the flat space ansatze, we use
\begin{eqnarray}
ds^2&=&\beta(y)^2\eta_{\mu\nu}dx^\mu dx^\nu +dy^2, \nonumber \\
H_{\mu\nu\rho\sigma}&=&\epsilon_{\mu\nu\rho\sigma}\frac{\sqrt{-g}}{n(y)},
\ \ H_{5\mu\nu\rho}=0.\label{ansatz1}
\end{eqnarray}
The $H$ field equation is
$\partial_M[\sqrt{-g}H^{MNPQ}/H^4]=\partial_\mu [\sqrt{-g}H^{\mu
NPQ}/H^4]=0$, and hence fixes $n$ as a function of $y$ only. The
two relevant Einstein equations are
\begin{eqnarray}
(55)\ &:&\
6\left(\frac{\beta^\prime}{\beta}\right)^2=-\Lambda_b-\frac{\beta^8}{A}
\nonumber\\
(\mu\nu)\ &:&\ 3\left(\frac{\beta^\prime}{\beta}\right)^2
+3\left(\frac{\beta^{\prime\prime}}{\beta}\right)= -\Lambda_b
-\Lambda_1\delta(y)-3\left(\frac{\beta^8}{A}\right)
\end{eqnarray}
where $\Lambda_b<0$ and $2n^2=\beta^8/A$ with $A>0$. We require
the $Z_2$ symmetry, and the bulk equation is easily solved. The
boundary condition at $y=0$ is
$(\beta^\prime/\beta)|_{0^+}=-\Lambda_1/6$. Then, we find a
solution
\begin{equation}
\beta(|y|)=\frac{1}{\left[(\frac{a}{k})\cosh(4k|y|+c)\right]^{1/4}}
\label{solution}
\end{equation}
where
\begin{equation}
k=\sqrt{\frac{-\Lambda_b}{6}},\ \ a=\sqrt{\frac{1}{6A}},\ \
k_1=\frac{\Lambda_1}{6}.
\end{equation}
The flat solution is shown in Fig. 3.

This solution has the integration constants $a$ and $c$. $a$
is basically the charge of the universe and determines the 4D
Planck mass. $c$ is the UIC which is fixed by the boundary
condition at $y=0$,
\begin{equation}
\tanh(c)
=\frac{k_1}{k}=\frac{\Lambda_1}{\sqrt{-6\Lambda_b}}.\label{uic}
\end{equation}
This solution shows that, for any value of $\Lambda_1$ in the
finite region allowing $\tanh$, it is possible to have a flat
space solution. Even if the observable sector adds some constant
to $\Lambda_1$, still it is possible to have the flat space
solution, just by changing the shape a little bit via $c$. The
change is acceptable since $H$ is a dynamical field. Note that
$\beta(y)$ is a decreasing function of $|y|$ in the large $|y|$
region and it goes to zero exponentially as
$|y|\rightarrow\infty$. We propose that this property is needed
for a self-tuning solution.

The key points found in our solution are\\
{\bf (i) $\beta$ has no singularity:} Our solution extends to
infinity without singularity, and $\beta^\prime\rightarrow 0$ as
$y\rightarrow\infty$.\\
{\bf (ii) 4D Planck mass is finite:} Even if the extra dimension
is not compact, this theory can describe an effective 4D theory
since gravity is localized. Integrating with respect to $y$, we
obtain an effective 4D Planck mass which is finite
\begin{eqnarray}
M^2_{4D\ Planck}&=&\int_{-\infty}^\infty dy \beta^2 \nonumber \\
&=& 2M^3\frac{k}{a}\int_0^\infty \frac{1}{[\cosh(4ky+c)]^{1/2}}dy
\end{eqnarray}
and is expected to be of order the fundamental parameters.\\
{\bf (iii) Self-tuning:} We obtained a self-tuning solution. To
check that the 4D c.c. is zero we integrate out the solution. For
that we have to include the surface term also
\begin{equation}
S_{surface}=\int d^4x dy 4\cdot
4!\partial_M\bigg[\sqrt{-g}\frac{H^{MNPQ}A_{NPQ}}{H^4}\bigg].\label{surface}
\end{equation}
Then the action is
\begin{eqnarray}
S&=&\int d^4x dy\sqrt{-g_4}\beta^4\bigg[\frac{1}{2\beta^2}R_4
-4\frac{\beta^{\prime\prime}}{\beta}-6
\left(\frac{\beta^\prime}{\beta}\right)^2 \nonumber \\
&-&\Lambda_b+\frac{2\cdot
4! }{H^2}-\Lambda_1\delta(y)\bigg]+S_{surface}.
\end{eqnarray}
Then, the effective 4D c.c. term $-\Lambda_{4D}$ is the $y$
integral except the $R_4$ term. One can show that $\Lambda_{4D}=0$
and it is consistent with the original ansatz of the flat
space\cite{kklcc0}.

\subsection{De Sitter and anti de Sitter space solutions}

For the de Sitter and anti de Sitter space solutions, the metric
is assumed as
\begin{eqnarray}
&ds^2=\beta(y)^2 g_{\mu\nu}dx^\mu dx^\nu +dy^2\nonumber\\
&g_{\mu\nu}=diag.\left(-1,e^{2\sqrt{\lambda}t},e^{2\sqrt{\lambda}t},
e^{2\sqrt{\lambda}t}\right),(dS_4, \lambda>0)\nonumber\\
&g_{\mu\nu}=diag.\left(-e^{2\sqrt{-\lambda}x_3},e^{2\sqrt{-\lambda}x_3},
e^{2\sqrt{-\lambda}x_3},1\right),(AdS_4, \lambda<0).
\end{eqnarray}
Note that $k=\sqrt{-\Lambda_b/6},k_1=\Lambda_1/6$, and the 4D
Riemann tensor is $R_{\mu\nu}=3\lambda g_{\mu\nu}$. The (55) and
(00) components equations are
\begin{eqnarray}
&6\left(\frac{\beta^\prime}{\beta}\right)^2-6\lambda\frac{1}{\beta^{2}}
=-\Lambda_b-3\frac{\beta^8}{A}\nonumber\\
&3\left(\frac{\beta^\prime}{\beta}\right)^2+3
\frac{\beta^{\prime\prime}}{\beta}-3\lambda\frac{1}{\beta^2}
=-\Lambda_b-\Lambda_1\delta(y)-3\frac{\beta^8}{A}.
\end{eqnarray}
The 4D c.c. obtained from the above ansatze is $\lambda$. Since we
cannot obtain the solution in closed forms, we cannot show this by
integration. However, we have checked this kind of
behavior\cite{kklcc} in the RS II model, using the Karch-Randall
form\cite{kr}. Here, we show just that the de Sitter and anti de
Sitter space solutions exist, and show the warp factor
numerically. In our model, the $y$ derivative of the metric is
\begin{equation}
\beta^\prime =\pm (k_\lambda^2+k^2\beta^2-a^2\beta^{10})^{1/2},\ \
k_\lambda=(\lambda/6)^{1/2}
\end{equation}
At $y=y_h$ where $\beta(y_h)=0$, $\beta^\prime(y_h)$ needs not be
zero due to the presence of the nonvanishing $k_\lambda$.
Therefore, there exists a point $y_h$ where $\beta^\prime$ is
finite. It is the de Sitter space horizon. It takes an infinite
amount of time to reach $y_h$. Also, we can see that it is
possible that $\beta$ can be nonzero where $\beta^\prime$ is zero
for $k_\lambda^2<0$. It is the anti de Sitter space solution.
These de Sitter space and anti de Sitter space solutions are shown
in Figs. 4 and 5. For the de Sitter space, we can integrate from
$-y_h$ to $+y_h$. As in the Karch-Randall example, it should give
the 4D c.c. $\lambda$. The AdS solution does not give a localized
gravity.

In the presence of de Sitter and anti de Sitter space solutions,
the c.c. problem relies on the old self-tuning solution. Namely,
the c.c. is probably zero, following Hawking\cite{hawking}.

Hawking showed that in 4D the probability is maximum, using the
Euclidian space action
\begin{equation}
-S_E=-\frac{1}{2}\int
d^4x\sqrt{+g}[(R+2\Lambda)+(1/48)H_{\mu\nu\rho\sigma}
H^{\mu\nu\rho\sigma}]
\end{equation}
where we use the unit $8\pi G=1$. The Einstein equation and field
equations are
\begin{eqnarray}
&R_{\mu\nu}-(1/2)g_{\mu\nu}R=\Lambda
g_{\mu\nu}-T_{\mu\nu}\nonumber\\
&\partial_\mu[\sqrt{g}H^{\mu\nu\rho\sigma}]=0.\nonumber
\end{eqnarray} From
the field equation, one has
$H^{\mu\nu\rho\sigma}=(1/\sqrt{g})\epsilon^{\mu\nu\rho\sigma} c$
or $H^2=4!c^2$. Thus, he obtained
\begin{eqnarray}
T^{\mu\nu}&=&-\Lambda_H g^{\mu\nu}, \Lambda_H=-c^2/2, \nonumber \\
R&=&-4\Lambda_{eff}, \Lambda_{eff}=\Lambda+\Lambda_H.
\end{eqnarray}
Thus,
\begin{eqnarray}
-S_E&=&-(1/2)\int d^4x\sqrt{g}(R+2\Lambda_{eff}) \nonumber \\
&=&{\rm Volume}\cdot
(\Lambda+\Lambda_H)=\frac{3M^4}{\Lambda_{eff}}
\end{eqnarray}
which is maximum at $\Lambda_{eff}=0^+$ which is shown in Fig. 6.

Note that Hawking used equations of motion.
Duff, on the other hand, used the action itself to calculate the
Euclidian action, and obtained
$-S_E=3M^4(\Lambda-(3/2)c^2)/(\Lambda-(1/2)c^2)^2$ which is
minimum at $0^+$\cite{duff}. But the consideration of the surface
term in the action would give additional contribution and should
give Hawking's result. The surface term is essential as we have
shown in our self-tuning solution.

Thus, the maximum probability occurs when the c.c. is zero. Our
self-tuning solution relies on this probabilistic choice of the
flat one from the flat, de Sitter and anti de Sitter space
solutions.

\section{Self-tuning solution in the dual description}

Let us now proceed to consider a dual description of the
self-tuning solution with $1/H^2$. When we introduce the $1/H^2$
term in the Lagrangian, the equation of motion for a three form
field A with H=dA is\footnote{The text mode(Roman) characters are
the form notation while equation mode(Italic) characters are the
component notation.}
\begin{equation}
{\rm d}\bigg(\frac{^*{\rm H}}{(H^2)^2}\bigg)=0
\end{equation}
where $^*$H is the Hodge dual of H. Note that we have not
introduced a source term for $A_{MNP}$.

The corresponding field strength H=dA should satisfy the Bianchi
identity, dH=0, locally. In the dual description, the roles of the
equation of motion and the Bianchi identity are interchanged from
the original one. Therefore, we can take a dual of the field
strength $H$ in 5D such that the equation of motion for $A$ is
trivially satisfied, as a Bianchi identity in the dual
description, as follows:
\begin{equation}
{\rm d}\sigma=\frac{^*{\rm H}}{(H^2)^2}\label{dual}
\end{equation}
where $\sigma$ is a dual scalar field. Then, we can identify the
field strength $H$ in terms of the scalar field $\sigma$ as
\begin{equation}
H_{MNPQ}=\sqrt{-g}\epsilon_{MNPQR}\frac{\partial^R\sigma}
{(4!)^{1/3}[(\partial\sigma)^2]^{2/3}}\label{dualcp}.
\end{equation}
As a result, the original Bianchi identity, dH=0, becomes in the
dual picture
\begin{equation}
\partial_M\bigg(\sqrt{-g}\frac{\partial^M\sigma}
{[(\partial\sigma)^2]^{2/3}}\bigg)=0,\label{BI}.
\end{equation}
which will be the equation of motion in the dual picture. In this
case, we can add a source term on the right hand side of Eq.
(\ref{BI}).

To show the self-tuning solution with $1/H^2$ in the dual picture
explicitly, let us insert the dual relation (\ref{dual}) or
(\ref{dualcp}) back to the Lagrangian and make a variation of the
action with respect to the dual scalar field . When the surface
term for $A$ is included, the dual 5D action is
\begin{equation}
S=\int d^4 x dy\,\sqrt{-g}\bigg(\frac{1}{2}R-
\gamma[(\partial\sigma)^2]^{1/3}-\Lambda_b-\Lambda_1
f(\sigma)\delta(y) \bigg) \label{daction}
\end{equation}
where $\gamma=6(4!)^{4/3}$ and $f(\sigma)$ is an arbitrary
coupling of the dual field to the brane, which breaks the shift
symmetry of the dual field at the brane.

Then, the energy-momentum tensor from the dual scalar field is
\begin{equation}
T^\sigma_{MN}=2\gamma\bigg(\frac{1}{3}[(\partial\sigma)^2]^{-2/3}
\partial_M\sigma\partial_N\sigma-\frac{1}{2}g_{MN}
[(\partial\sigma)^2]^{1/3}\bigg)\label{em}.
\end{equation}
And the equation of motion for the dual scalar field becomes
\begin{equation}
\partial_M\bigg(\sqrt{-g}\frac{\partial^M\sigma}
{[(\partial\sigma)^2]^{2/3}}\bigg)=\sqrt{-g^{(4)}}\Lambda_1
\frac{df}{d\sigma}\delta(y)
\end{equation}
where the original Bianchi identity (\ref{BI}) is not respected at
the brane due to the coupling $f(\sigma)$, i.e.,
\begin{equation}
dH_{5\mu\nu\rho\sigma}=\frac{1}{(4!)^{1/3}}\sqrt{-g^{(4)}}
\epsilon_{\mu\nu\rho\sigma}\Lambda_1\frac{df}{d\sigma}\delta(y).
\end{equation}
Note that the self-tuning solution with $1/H^2$ is the case with
$\frac{df}{d\sigma}$=0. In the dual picture, a new Bianchi
identity is dd$\sigma=0.$ Thus, the theory we study is different
from the original one in that the bulk field $\sigma$ couples to
the brane with $f(\sigma)$.

If we take the ansatze for the metric and the scalar field as
\begin{equation}
ds^2=\beta^2(y)\eta_{\mu\nu}dx^\mu dx^\nu+dy^2,~~\sigma=\sigma(y),
\end{equation}
then the relevant Einstein's equations and the scalar equation are
\begin{eqnarray}
3\bigg(\frac{\beta'}{\beta}\bigg)^2+3\frac{\beta^{\prime\prime}}{\beta}
&=&-\Lambda_b-\Lambda_1 f(\sigma)\delta(y)-\gamma |\sigma'|^{2/3},
\label{dmumu}\\
6\bigg(\frac{\beta'}{\beta}\bigg)^2&=&-\Lambda_b-\frac{1}{3}\gamma
|\sigma'|^{2/3}, \label{d55}\\
\bigg(\beta^4\frac{\sigma'}{|\sigma'|^{4/3}}\bigg)'
&=&\beta^4\Lambda_1\frac{df}{d\sigma}\delta(y)\label{dscalar}.
\end{eqnarray}
Thus, we find that the solutions of the above three equations
consistent with the $Z_2$ symmetry become
\begin{eqnarray}
\beta(y)&=&\bigg(\frac{k}{a}\bigg)^{1/4}[{\rm
cosh}(4k|y|+c)]^{-1/4},
\label{warp}\\
\sigma'(y)&=&\pm\bigg(\frac{18a^2}{\gamma}\bigg)^{3/2}
\beta^{12}\epsilon(y)\label{sigma}
\end{eqnarray}
where $\epsilon(y)$ is the step function which is +1 for $y>0$ and
--1 for $y<0$, and
\begin{equation}
k=\sqrt{-\frac{\Lambda_b}{6}}, ~~a=\sqrt{\frac{1}{6A}}.
\end{equation}
On the other hand, from Eqs.~(\ref{dmumu}) and (\ref{dscalar}),
the boundary conditions at the brane for the metric and the scalar
are given by
\begin{eqnarray}
\frac{\beta'}{\beta}\Big|_{y=0^+}&=&-k_1 f(\sigma(0)),
\label{mbc}\\
\frac{\sigma'}{|\sigma'|^{4/3}}\Big|_{y=0^+}
&=&3k_1\frac{df}{d\sigma}(\sigma(0))\label{sbc}
\end{eqnarray}
where
\begin{equation}
k_1=\frac{\Lambda_1}{6}.
\end{equation}
Therefore, there arise two consistency conditions due to the
existence of the brane and the scalar coupling :
\begin{eqnarray}
{\rm tanh}(c)&=&\frac{k_1}{k}f(\sigma(0)),\label{cond1}\\
\pm\sqrt{\frac{\gamma}{2}}\frac{1}{k}{\rm cosh}(c) &=&9k_1
\frac{df}{d\sigma}(\sigma(0))\label{cond2}.
\end{eqnarray}
That is, the condition for the scalar coupling function at the
brane becomes
\begin{equation}
\frac{df}{d\sigma}(\sigma(0))=\pm\sqrt{\frac{\gamma}{162}}\frac{1}{k
k_1} \bigg(1-\bigg(\frac{k_1}{k}\bigg)^2
f^2(\sigma(0))\bigg)^{-1/2} \label{coupling}.
\end{equation}
Without the scalar coupling, Eq.~(\ref{cond2}) is absent and
Eq.~(\ref{cond1}) is the self-tuning solution obtained in terms of
the three form field\cite{kklcc0}, and we reproduce the solution
in the dual picture.

Even with the scalar coupling, i.e. the original Bianchi identity
becoming the equation of motion in the dual picture with the
scalar coupling to the brane, the self-tuning property is
maintained because $\sigma(0)$ acts as an integration constant of
Eq.~(\ref{sigma}). For a given set of $\Lambda_b,\Lambda_1,$ and
parameters appearing in $f(\sigma)$, Eq. (\ref{coupling}) can be
satisfied by choosing an appropriate value of the integration
constant $\sigma(0)$. For example, if $f(\sigma)=e^{b\sigma}$ then
for some $b$ we can find $\sigma(0)$ as far as Eq.
(\ref{coupling}) is satisfied. Note that the condition Eq.
(\ref{coupling}) is given only for the field value at $y=0$. Of
course, there also exist de Sitter and anti de Sitter space
solutions. Thus, our self-tuning solution is of the old style.

To see it explicitly, let us consider $f(\sigma)$ as an exponential function
\begin{equation}
f(\sigma)=e^{b\sigma}
\end{equation}
where we set the overall constant as 1 and $b$ is a number. Then, 
Eq. (\ref{coupling}) becomes
\begin{equation}
b^2e^{2b\sigma(0)}\left[1-\frac{k_1^2}{k^2}e^{2b\sigma(0)}
\right]=\frac{\gamma}{162}\frac{1}{(kk_1)^2}
\end{equation}
which is a quadratic equation of $e^{2b\sigma(0)}$. For a finite range
of parameters, it has two solutions, proving the self-tuning property.
It is different from the Kachru et al. solution which required
a fine-tuning of $b$ as $4/3$.

\section{conclusion}

It is pointed out that a self-tuning solution of the cosmological
constant is possible in the RS II type models. We have shown this
$1/H^2$ term in the Lagrangian as a review to guide the
self-tuning solution\cite{kklcc0,kklcc}. We presented this
solution in the dual picture. The kinetic energy term in the dual
picture has a peculiar form $((\partial_\mu\sigma)^2)^{1/3}$. In
the dual picture, this self-tuning property is still maintained
even with the presence of the dual field coupling to the brane.
However, the new type of self-tuning solution is not obtained. The
self-tuning relies on the old idea of choosing the flat one out of
numerous possibilities\cite{hawking}.

The dual picture description is intriguing. As discussed in Sec.
IV, the scalar field $\sigma$ has an integration constant
$\sigma(0)$. We can choose $\sigma(0)$ such that a flat space
solution results. However, for another nearby value of
$\sigma(0)$, a de Sitter or anti de Sitter space solution is
possible, which is the reason that we could have not obtained a
new type of self-tuning solution. The unique flat space solution
requires a fine-tuning of $\sigma(0)$, at present. In the future,
it will be interesting to see if some principle chose such a
fine-tuning of $\sigma(0)$.

\acknowledgments This work is supported in part by the BK21
program of Ministry of Education, the KOSEF Sundo Grant, by the
Center for High Energy Physics(CHEP), Kyungpook National
University, and CTP Support Grant of SNU.

\newpage
\begin{figure}[ht]
\centering
\centerline{\epsfig{file=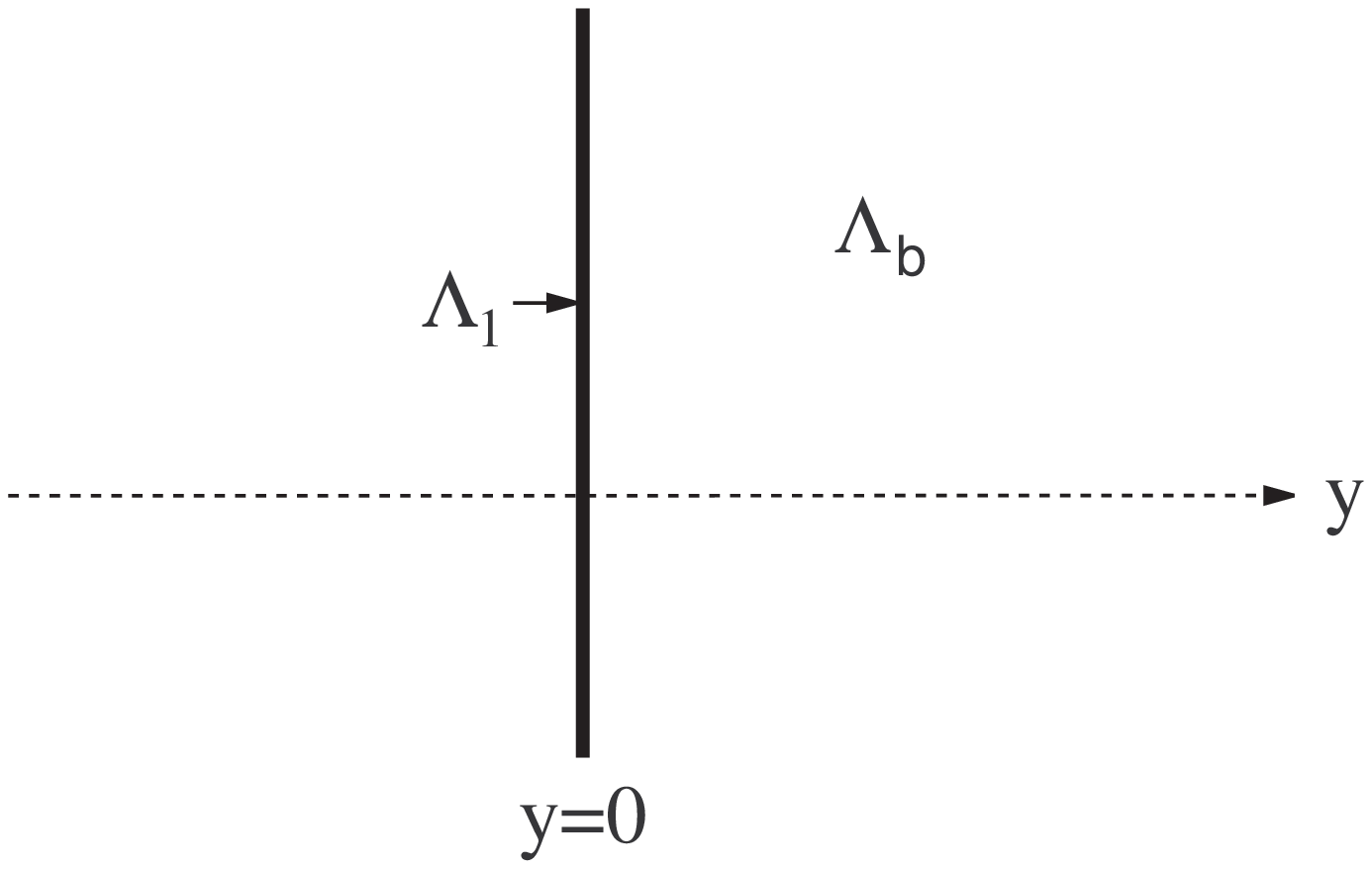,width=65mm}}
\end{figure}
\centerline{ Fig.~1.\  The RS II model with a
brane at $y=0$. 
}
\vskip 1.5cm

\begin{figure}[ht]
\centering
\centerline{\epsfig{file=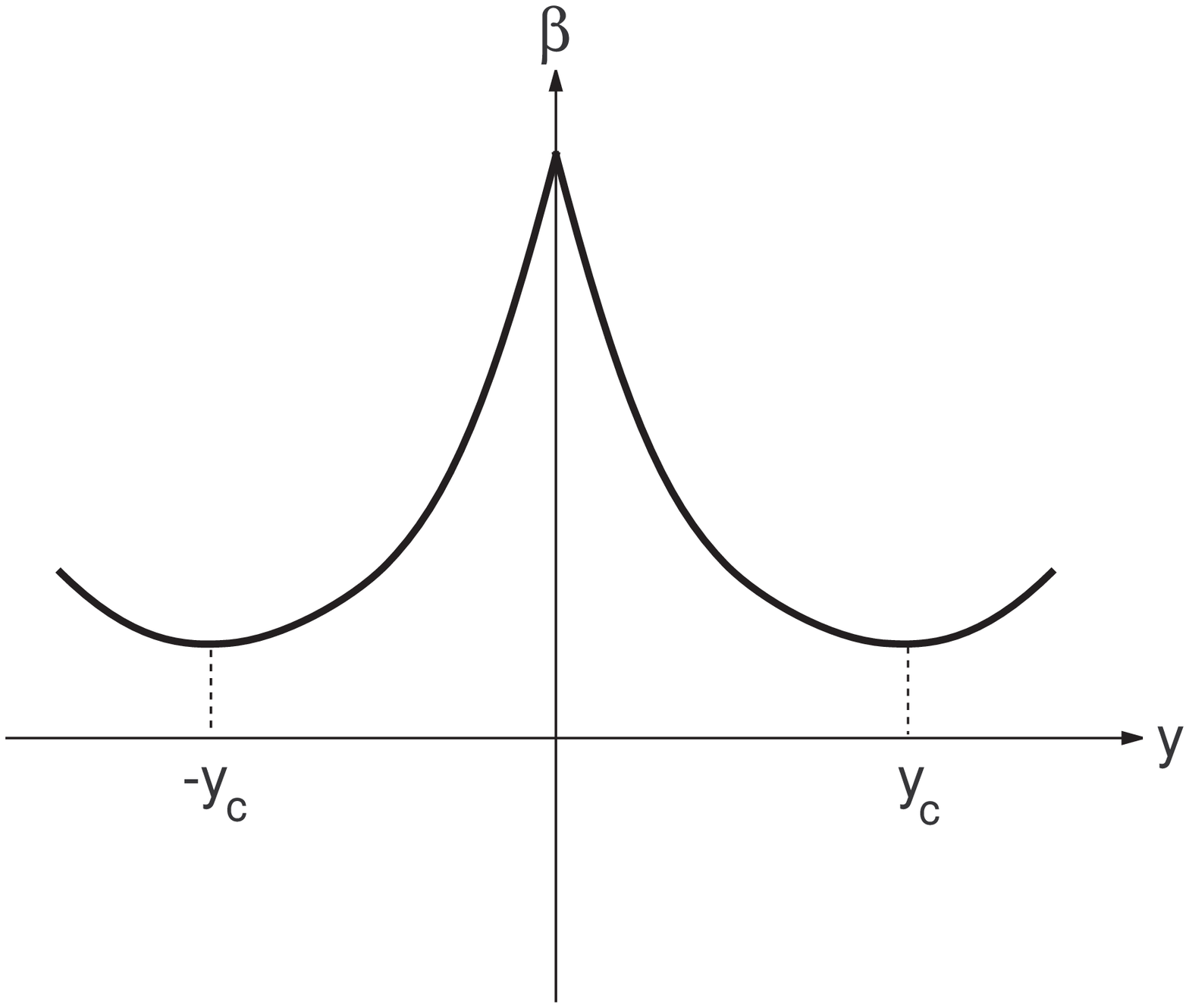,width=65mm}}
\end{figure}
\centerline{ Fig.~2.\  The flat space solution with $H^2$
}
\vskip 1.5cm

\begin{figure}[ht]
\centering
\centerline{\epsfig{file=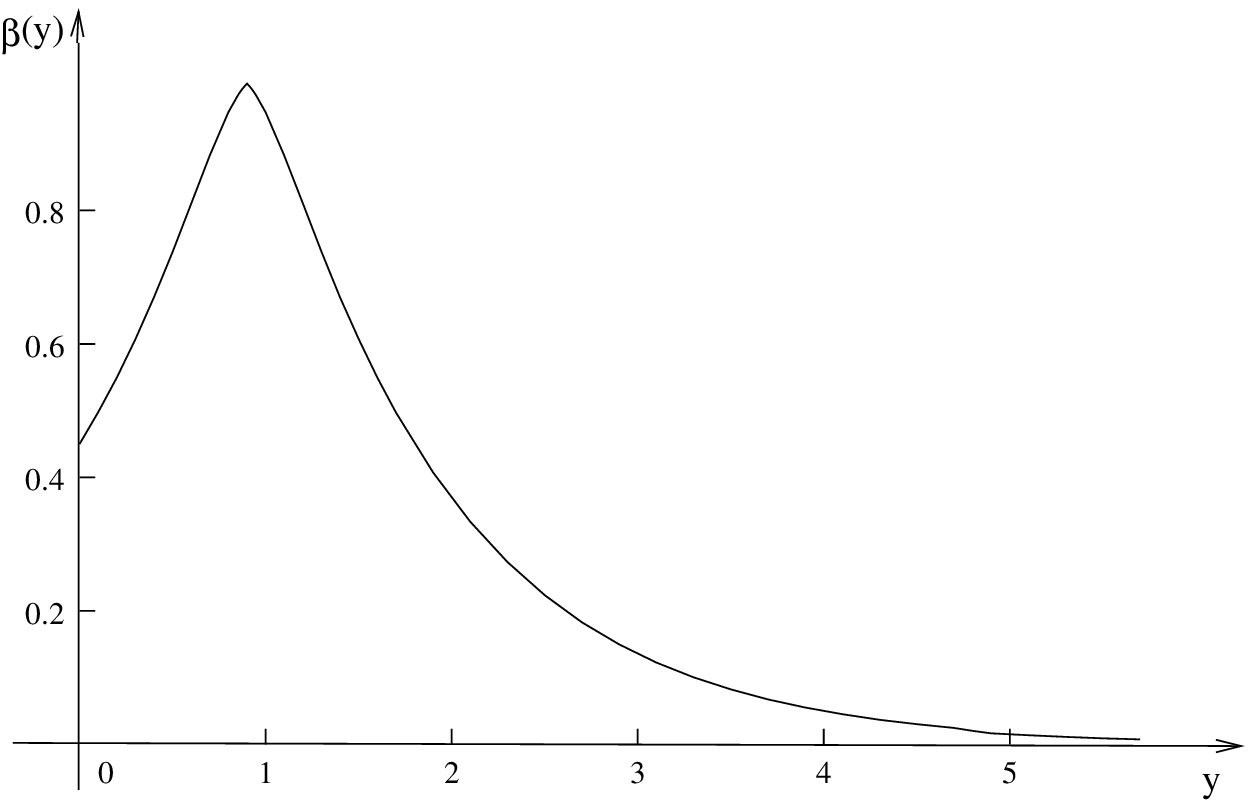,width=50mm}}
\end{figure}
\centerline{ Fig.~3.\  The flat space solution.
}

\newpage
\begin{figure}[ht]
\centering
\centerline{\epsfig{file=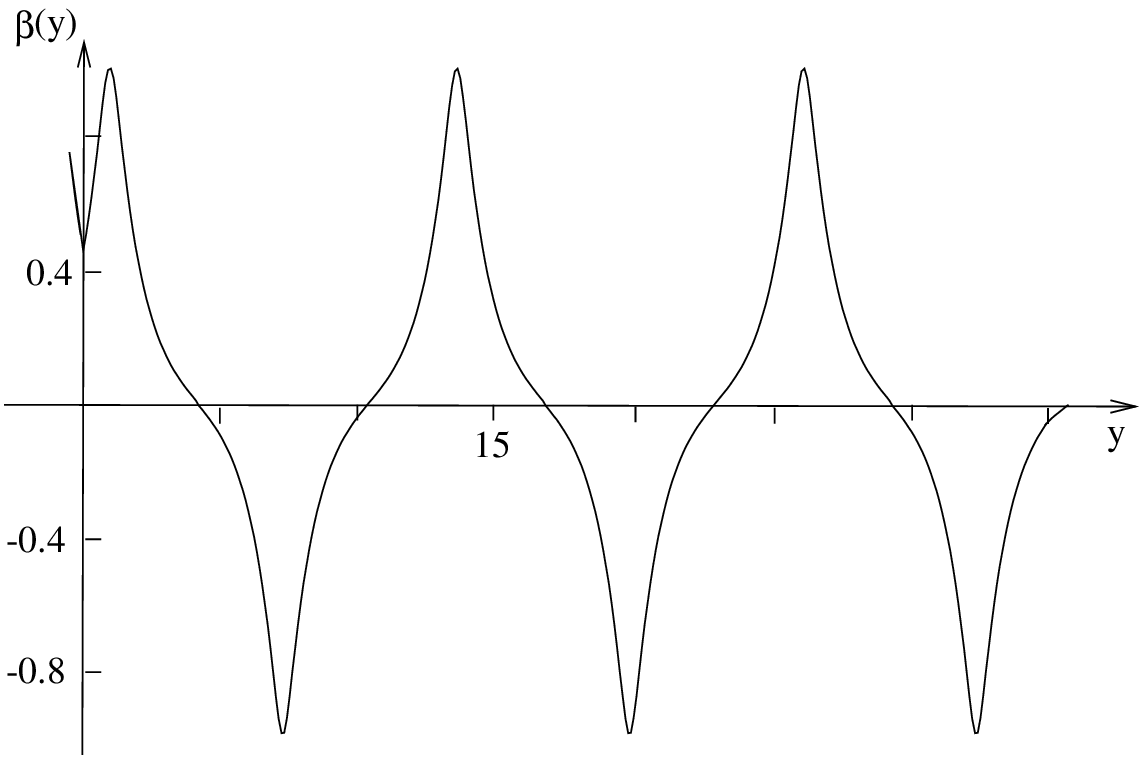,width=50mm}}
\end{figure}
\centerline{ Fig.~4.\  The de Sitter space solution.
}
\vskip 1.5cm

\begin{figure}[ht]
\centering
\centerline{\epsfig{file=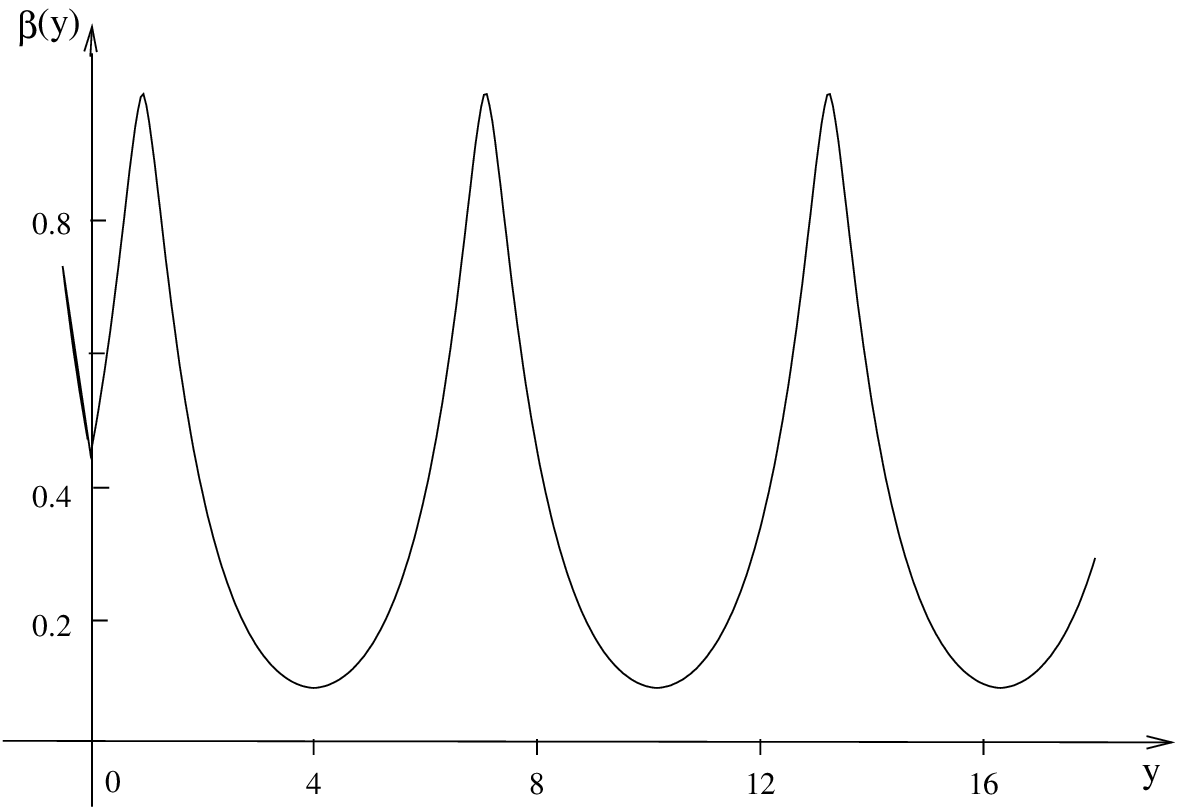,width=50mm}}
\end{figure}
\centerline{ Fig.~5.\  The anti de Sitter space solution.
}
\vskip 1.5cm

\begin{figure}[ht]
\centering
\centerline{\epsfig{file=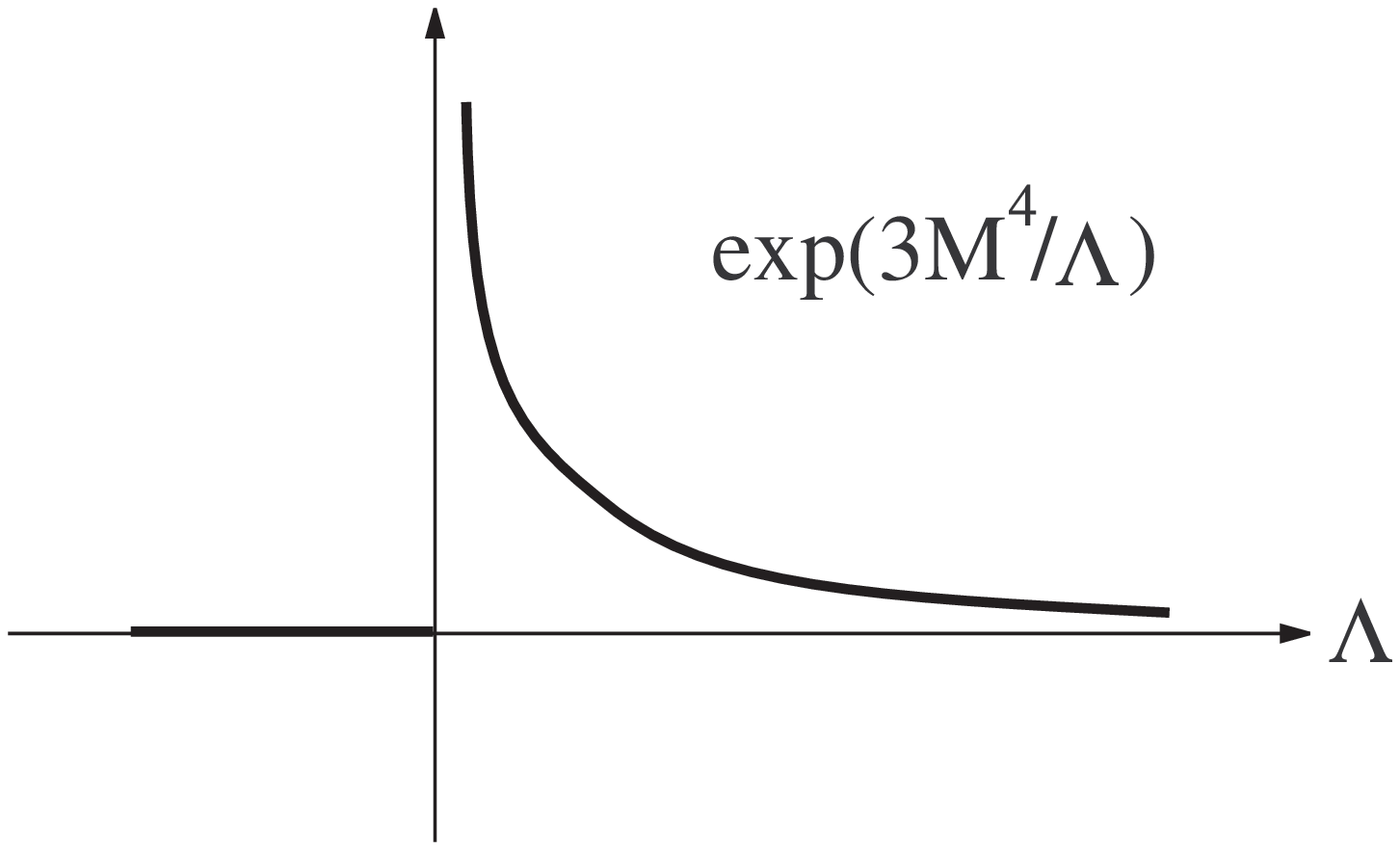,width=65mm}}
\end{figure}
\centerline{ Fig.~6.\  Hawking's probability
}

\end{document}